\documentclass[aps,pra,
               twocolumn,superscriptaddress,
               floatfix,
               longbibliography, 
               showkeys]{revtex4}  
  
\usepackage{graphicx} 
\usepackage{bm}	
\usepackage{color}                      
\usepackage{xcolor}
\usepackage{epsfig} 
\usepackage{amsmath} 
\usepackage{amssymb} 
\usepackage{longtable}  
\usepackage{float}
\usepackage{mathtools}
\usepackage{xparse}
\usepackage{hyperref} 
\usepackage[normalem]{ulem}

\usepackage{qcircuit}

\newtheorem{thm}{Theorem}

\smallskip
\def\<{{\langle }}
\def\>{{\rangle }}

\def\ket#1{|#1\rangle}

\smallskip
\def\<{{\langle }}
\def\>{{\rangle }}

\begin{document}
\title{ Entanglement types for two-qubit states with real amplitudes}

\author{Oscar Perdomo}
\affiliation{Rigetti Computing, 2919 Seventh Street, Berkeley, CA 94710-2704, USA}
\affiliation{Department of Mathematics, Central Connecticut State University, New Britain, CT 06050, USA}

\author{Vicente Leyton-Ortega}
\affiliation{Computer Science and Engineering Division, Oak Ridge National Laboratory,
One Bethel Valley Road, Oak Ridge, TN 37831 USA}
\affiliation{Rigetti Computing, 2919 Seventh Street, Berkeley, CA 94710-2704, USA}

\author{Alejandro Perdomo-Ortiz}
\email{alejandro@zapatacomputing.com}
\affiliation{Zapata Computing Canada Inc., 1 Yonge Street, Suite 900, Toronto, ON, M5E 1E5}
\affiliation{Department of Computer Science, University College London, WC1E 6BT London, UK}
\affiliation{Rigetti Computing, 2919 Seventh Street, Berkeley, CA 94710-2704, USA}

\date{\today}  

\begin{abstract}
    We study the set of two-qubit pure states with real amplitudes and their geometrical representation in the three-dimensional sphere. In this representation, we show that the maximally entangled states --those locally equivalent to the Bell States --form two disjoint circles perpendicular to each other. We also show that taking the natural Riemannian metric on the sphere, the set of  states connected by local gates are equidistant to this pair of circles. Moreover, the unentangled, or so called product states, are $\pi/4$ units away to the maximally entangled states. This is, the unentangled states are the farthest away to the maximally entangled states. In this way, if we define two states to be equivalent if they are connected by local gates, we have that there are as many equivalent classes as points in the interval $[0,\pi/4]$ with the point $0$ corresponding to the maximally entangled states. The point $\pi/4$ corresponds to the unentangled states which geometrically are described by a torus. Finally, for every $0<  d < \pi/4$ the point $d$ corresponds to a disjoint pair of torus. We also show that if a state is $d$ units away from the maximally entangled states, then its entanglement entropy is $S(d)  = 1- \log_2 \sqrt{\frac{(1+\sin 2 d)^{1+\sin 2 d}}{(1-\sin 2 d)^{-1+\sin 2 d}}}$.  Finally, we also show how this geometrical interpretation allows to clearly see that any pair of two-qubit states with real amplitudes can be connected with a circuit that only has single-qubit gates and one controlled-Z gate.
\end{abstract}



\maketitle

\section{Introduction}

Entanglement is one of the fascinating properties of multipartite quantum systems, and one of the cornerstones towards the existence of many quantum technologies; from the design of quantum communication protocols (e.g., quantum teleportation and quantum cryptography) to the success of many key quantum computer algorithms. The geometry of entanglement, and in general, the geometry of quantum states, has been invaluable and extensively studied by many authors in the past decades \cite{Brody2001,bengtsson2006,Jevtic2014,Milne2014,Boyer2017,Avron2019}. In this work, we geometrically describe how the space of pure states of 2-qubit systems with real amplitudes partitions when we group them by their entanglement. 

In order to study the set of 2-qubit states with real amplitudes we organize them into ``orbits'' using the following condition: two 2-qubit states are in the same orbit if they are connected by local gates with real entries, this is, if they are connected by a circuit that uses only the set of gates $L^{\mathbb{R}} = \{R_y(\theta), X\}$ with $R_y(\theta)=\begin{pmatrix} \cos(\theta/2) &-\sin(\theta/2)\\ \sin(\theta/2) &\cos(\theta/2) \end{pmatrix}$  and the $X=\begin{pmatrix} 0 &1\\ 1 &0 \end{pmatrix}$ gates. We will refer to gates generated by $L^{\mathbb{R}}$ as real local gates.
The whole set of 2-qubits with real amplitudes is described by the three dimensional unit sphere and, for any 2-qubit state, the orbit that it belongs to is either a pair of disjoint circles, a pair of disjoint tori or a single torus. We will also show that, with respect to the natural Riemannian metric of the 3-dimensional sphere, these orbits are described as the set of points that are at the same distance $d$ from the only orbit that consist of two disjoint circles. 

The equivalent relation considered here is not new. For the general case of $n$-qubits with complex amplitudes, the authors \cite{Walck2005,Znidaric2008,Linden1998} have considered the equivalence relation: Two $n$-qubit states are equivalent if they are connected by local gates. Since they are considering $n$-qubits with complex amplitudes there are not restrictions on the local gates, i.e., gates of the form $U_1\otimes\dots\otimes U_n$ where each $U_i$ is a $2\times 2$ unitary matrix. Each orbit is called an ``entanglement type'' and the quotient space is called "the space of entanglement types''; the latter denoted as $\mathcal{E}_n$. In \cite{Walck2005} the author shows that $\mathcal{E}_2$ is a closed interval and the orbits are either a 4-dimensional manifold that contains the unentangled 2-qubit (product) states or, a 3-dimensional manifold that contains the EPR (maximally entangled) states or, 5-dimensional manifolds for all other cases with intermediate entanglement strength. Placing our result in the same perspective, let us denote by $\mathcal{E}^{\mathbb{R}}_2$ the quotient space of 2-qubit states with real amplitudes where two 2-qubit states are identified if they can be connected with real local gates. The space $\mathcal{E}^{\mathbb{R}}_2$ is also a closed interval, the interval $[0,\pi/4]$, and the orbits are either: (i) a connected 2-dimensional manifold -a torus- that contains all the unentangled 2-qubits product states with real amplitudes, (ii) a pair of disjoint circles (1-dimensional manifolds) that contains the EPR maximally entangled states, or (iii) a pair of tori (2-dimensional manifold) representing all other states with intermediate entanglement values. 

We would like to point out that our result is not a consequence of Walck's \cite{Walck2005} result due to the fact that both, the spaces and the equivalent relations involved are different. In fact, one can easily show with arguments based on the dimension of the manifolds involved that the entanglement spaces $\mathcal{E}_n$ and $\mathcal{E}^{\mathbb{R}}_n$ are different for all $n>2$.  An example related with this difference in given in \cite{Oscar3qubits}, where it is shown a pair of 3-qubit states with real amplitudes that are part of the same orbit in $\mathcal{E}_3$ but that are in two different orbit in $\mathcal{E}^{\mathbb{R}}_3$. 

Finally, we show that when we apply the controlled-$Z$ gate (CZ) to a particular equivalent class, we get a set containing points from all the equivalent classes. As a consequence, this provides a geometrical proof that every pair of real 2-qubit states can be connected by a circuit that contains local gates and only one CZ gate. That only one entangling gate is necessary for arbitrary 2-qubit states is a known result and it fits within the general theory and research devoted to efficiently prepare $n$-qubit states (see e.g., ~\cite{Shende2006,Plesch2011,Iten2016}, and for a subroutine in Rigetti's Quil platform, see~\cite{Quil}).

Section \ref{s:geo} introduces a geometrical interpretation of the orbits coming for the partition induced by $\mathcal{E}^{\mathbb{R}}_3$ and how they relate with each other. Section \ref{s:main} presents the proof of our main result and finally, section \ref{s:conclusions} shows the conclusions.

\section{Geometrical interpretation}\label{s:geo}

To describe the geometry of the orbits of 2-qubit states with real amplitudes, we will start by describing a similar situation one dimension lower. Let us assume that the Earth is a perfect sphere with radius 1, defined by $\mathbb{S}^2 = \lbrace \bm{x} \in \mathbb{R}^2 : || \bm{x} || = 1 \rbrace $ (with $|| \cdot ||$ as the Euclidean norm). In this case, the equatorial line is a unit circle. The name equatorial {\bf line} turns out to be appropriate because unit circles on the sphere are geodesics, this means if we consider two points on the Equator, the shortest path connecting these points along the surface of the Earth must be part of the equatorial line. Now, let us think of the points on Earth that are at a spherical distance $d$ from the equatorial line and let us denote this set as $\Sigma_d$. For $d = \pi / 4$, the set $\Sigma_{d}$ consists of all those points on the Earth that are on a latitude $45$ degrees north and $45$ degrees south. It is not difficult to see that $\Sigma_d$, for  $0 < d <\pi/2$, is the union of two disjoint circles of Euclidean radius $\cos(d)$. When $d=\pi/2$, $\Sigma_d$ reduces to only two points, the south and north pole.

For the study of two-qubits quantum states with real amplitudes, let us consider the three dimensional sphere defined by $\mathbb{S}^3= \lbrace \bm{x} \in \mathbb{R}^4 : ||\bm{x} || = 1 \rbrace$.  Given an orthonormal basis $\bm{v}_1$, $\bm{v}_2$, $\bm{v}_3$, $\bm{v}_4$ for $\mathbb{R}^4$, let us define the set $\, E(\bm{v}_3,\bm{v}_4)=\{\cos(\theta) \bm{v}_3+\sin(\theta) \bm{v}_4,\theta\in\mathbb{R}\}$. Clearly $E$ is a unit circle which is a geodesic in $\mathbb{S}^3$. For any positive $d<\pi/2$, let us consider the set $\Sigma_d$ of points in $\mathbb{S}^3$ that are exactly at a distance $d$ from the circle $E$. This time the set $\Sigma_d$ is not the disjoint union of two circles like in the case of the 2-dimensional sphere, it is just one torus.  More precisely, if $\bm{x}=x_1\bm{v}_1+x_2\bm{v}_2+x_3\bm{v}_3+x_4\bm{v}_4$,

\begin{eqnarray}
\Sigma_d(\bm{v}_3,\bm{v}_4)= \big\lbrace \bm{x} \in \mathbb{S}^3: x_1^2+x_2^2&=&\sin^2(d),\, \nonumber \\ 
 x_3^2+x_4^2 &=& \cos^2(d) \big\rbrace \ .
\end{eqnarray}

 A direct computation shows that, viewing $\Sigma_d(\bm{v}_3,\bm{v}_4)$ as a surface of $\mathbb{S}^3$, its Gauss curvature is zero and the principal curvatures are $\tan(d)$ and $-\cot(d)$. The surface $\Sigma_{\pi/4}$ played an important role in the study of minimal surface on $\mathbb{S}^3$ since it was conjectured by Lawson in 1970 \cite{Lawson1970} that this surface was the only embedded minimal torus on the sphere. The conjecture was finally solved by Brendle in 2013 \cite{Brendle2013}. When $d=\pi/2$ the set $\Sigma_d$ reduces to a circle of radius 1, $\, \Sigma_{\pi/2 }(\bm{v}_3,\bm{v}_4)=\{ \cos(\theta) \bm{v}_1+\sin(\theta) \bm{v}_2,\, \theta\in \mathbb{R} \}$. Notice that $\Sigma_{\pi/2 }(\bm{v}_3,\bm{v}_4)=E(\bm{v}_1,\bm{v}_2)$.

As shown in detail in section II,  a remarkably simple geometrical representation emerges when we perform the analysis above on the appropriate orthogonal basis $\bm{v}_1$, $\bm{v}_2$, $\bm{v}_3$, $\bm{v}_4$.

As part of the geometric interpretation that we are providing we need to identify 2-qubit states with real amplitudes with either points in $\mathbb{S}^3$ or with vectors in $\mathbb{R}^4$. For this reason, we may say ``the vector $\frac{1}{\sqrt{2}}|00> - \frac{1}{\sqrt{2}}|11>$", referring to the vector $\frac{1}{\sqrt{2}}(1,0,0,-1) $. In general we make the qubit $w_1 \ket{00}+ w2 \ket{01}+ w_3 \ket{10}+ w_4 \ket{11}$ indistinguishable from $(w_1,w_2,w_3,w_4)$. Selecting the following orthonormal basis

 \begin{eqnarray}\label{eq:BellBasis}
 \ket{v_1}&=&\frac{1}{\sqrt{2}}(\ket{00}-\ket{11}),\quad \ket{ v_2}=\frac{1}{\sqrt{2}}(\ket{01}+\ket{10}),\nonumber \\
  \ket{v_3}&=&\frac{1}{\sqrt{2}}(\ket{00}+\ket{11}),\quad \ket{v_4}=\frac{1}{\sqrt{2}}(\ket{01}-\ket{10}) , \nonumber \\  
 \end{eqnarray}

will allows us to describe the orbits that we obtain by placing together 2-qubits that can be connected by local gate with real entries. To start with, the Bell states \eqref{eq:BellBasis} are all in the same orbit. This orbit is  the union of the two circles $E(\bm{v}_3,\bm{v}_4)$ and $E(\bm{v}_1,\bm{v}_2)$. For any $0<d<\pi/4$, the orbit is the union of the two tori $\Sigma_d(\bm{v}_1,\bm{v}_2)$ and
$\Sigma_d(\bm{v}_3,\bm{v}_4)$ mades up an orbit. And when $d=\pi/4$, we have that the tori $\Sigma_d(\bm{v}_1,\bm{v}_2)$ and
$\Sigma_d(\bm{v}_3,\bm{v}_4)$ are the same. This torus $\Sigma=\Sigma_d(\bm{v}_1,\bm{v}_2)=\Sigma_d(\bm{v}_3,\bm{v}_4)$ mades up an orbit and it contains all the unentangle states. We can visualize half of the 3-dimensional sphere using the parametrization

\begin{equation}
  \xi(u)=\left(u_1,u_2,u_3,\sqrt{1-u_1^2-u_2^2-u_3^2} \right). 
\end{equation}

\begin{figure*}[t]
  \includegraphics[width=0.8\textwidth]{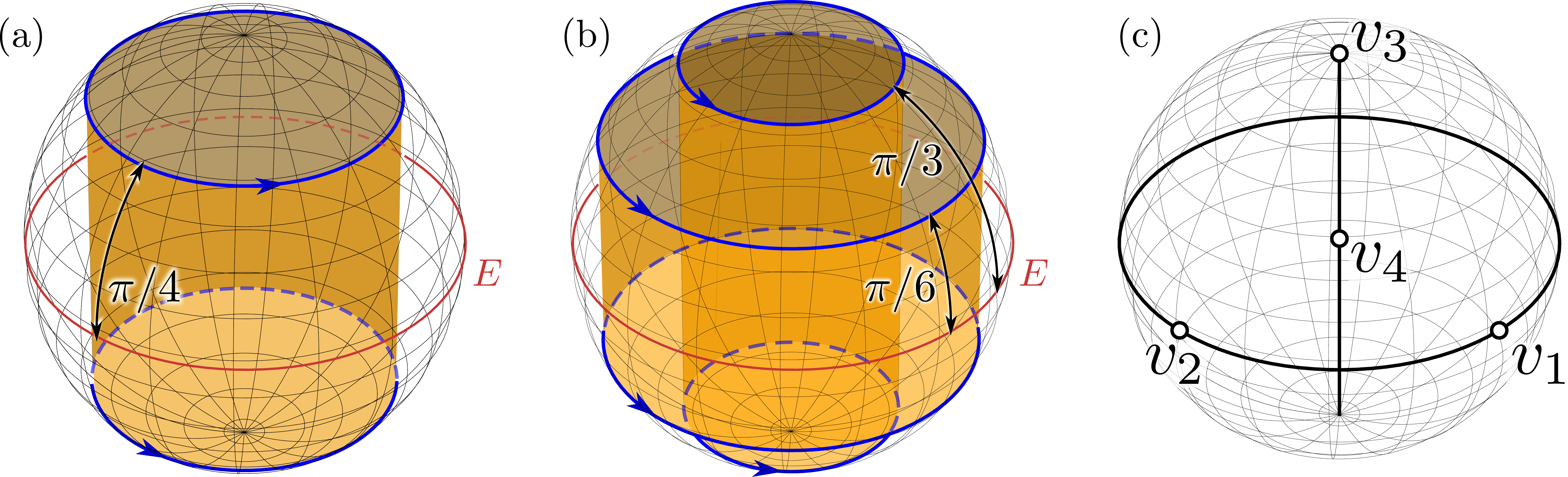}
  \caption{Half of the 3-dimensional sphere  projects into the unit ball in $\mathbb{R}^3$. This Figure shows the projection of the typical orbits on this ball.  (a)  Projection of the orbit that contains the  unentangled or product states. The whole orbit is a torus and this image shows half of it. (b) Projection of a generic orbit, here we show the one  associated with  $d=\pi/6$. For any state associated with $d$ between $0$ (Bell states) and $\pi/4$ (product states), its entanglement entropy $S(d) =  1- \log_2 \sqrt{\frac{(1+\sin 2 d)^{1+\sin 2 d}}{(1-\sin 2 d)^{-1+\sin 2 d}}}$ (c) Projection of the orbit that contains the Bell states. This Orbit is the union of two circles. Here we show one of the circles and half of the other circle}~\label{f:cases}
\end{figure*}

This parametrization identifies the unit ball in $\mathbb{R}^3$ with half of the sphere. We will show the projection on the unit ball of $\mathbb{R}^3$  of the part of the orbits that lies on the half part of the 3-dimensional sphere parametrized by $\xi$.  Fig.~\ref{f:cases}(a) shows the projection of the orbit that contains the unentangled states.Fig.~\ref{f:cases}(b) shows an orbit for a $d$ between $0$ and $\pi/4$, and finally Fig.~\ref{f:cases}(c) shows the orbit that contains the Bell states.

\section{Main theorem and proof} \label{s:main}

Let us recall that the action of the $\rm CZ$ gate on a pair of qubits,  is given by matrix

$$ \left(
\begin{array}{cccc}
 1 & 0 & 0 & 0 \\
 0 & 1 & 0 & 0 \\
 0 & 0 & 1 & 0 \\
 0 & 0 & 0 & -1 \\
\end{array}
\right) \ .$$ 

Up to local gates, the CNOT gate and the CZ gate are equivalent.  We are now ready to state and proof the main theorem in this paper.

\begin{thm} \label{thm1}
Let us consider the following subset of two qubits $  RQ_2$ given by

\begin{eqnarray}\label{2-qubit}
\{\ket{w}=w_1\ket{00}+ w_2\ket{01}+w_3\ket{10}+w_4\ket{11}: w_i\in\mathbb{R}\}
 \end{eqnarray}

 and let us define the Bell basis as
 
 \begin{eqnarray}
 \ket{v_1}&=\frac{1}{\sqrt{2}}(\ket{00}-\ket{11}), \quad 
 \ket{ v_2}&=\frac{1}{\sqrt{2}}(\ket{01}+\ket{10}),\nonumber \\
 \ket{v_3}&=\frac{1}{\sqrt{2}}(\ket{00}+\ket{11}), \quad \ket{v_4}&=\frac{1}{\sqrt{2}}(\ket{01}-\ket{10}), \nonumber 
 \end{eqnarray}
 
 and let us call $(x_1,x_2,x_3,x_4)$ the coordinates of $RQ_2$ with respect to the basis $\ket{v_1},\dots,\ket{ v_4}$. Notice that for the state $\ket{w}=w_1\ket{00}+ w_2\ket{01}+w_3\ket{10}+w_4\ket{11}$ the following relation holds,
 
 \begin{eqnarray}
 x_1 &= \dfrac{w_1-w_4}{\sqrt{2}} ,
 \quad x_2 & = \dfrac{w_2+w_3}{\sqrt{2}} \nonumber \\
  x_3 & = \dfrac{w_1+w_4}{\sqrt{2}},
  \quad x_4 & = \dfrac{w_2-w_3}{\sqrt{2}} \ . \nonumber 
  \end{eqnarray}
 
 Let us consider the equivalence relation $\ket{\psi_1}\sim\ket{\psi_2}$ if there is a local gate with real entries that connects them. We will refer to the orbit that contains a given  $\ket{\psi}$ to the set of 2-qubit states that are equivalent to $\ket{\psi}$, this is the set of 2-qubit states that can be connected with $\ket{\psi}$ by means of  local gates with real entries. Using the notation introduced earlier we have,

 \begin{enumerate}
 \item The orbit that contains the state $\ket{v_1}$ is the union of the two circles $E(\bm{v}_3,\bm{v}_4)$ and 
 $E(\bm{v}_1,\bm{v}_2)$. Therefore, the 2-qubit states $\ket{v_1}, \ket{v_2},\ket{v_1}$ and $\ket{v_4}$ are part of the same orbit.
 This orbit is the only one give by a 1-dimensional manifold.
 
 \item The orbit that contains the state $\ket{00}$ is the torus $\Sigma_{\pi/4}(\bm{v}_3,\bm{v}_4)=\Sigma_{\pi/4}(\bm{v}_1,\bm{v}_2)$. This orbit contains all the unentangled states. Notice that this set is also characterized as the set of point that are farthest away from the orbit that contains the Bell states. The distance between these two orbits is $\pi/4$.
 
 \item Any other orbit is given by the union of the two tori $\Omega_d=\Sigma_d(\bm{v}_1,\bm{v}_2)\cup \Sigma_d(\bm{v}_3,\bm{v}_4)$ where $d$ lies between $0$ and $\pi/4$. As a consequence,  Any pair  of states in $\Omega_d$ are connected by local gates and, if $0\le d_1<d_2\le \pi/4$ and $\ket{\phi_1}\in \Omega_{d_1}$ and $\ket{\phi_2}\in \Omega_{d_2}$ then, $\ket{\phi_1}$ and $\ket{\phi_2}$ are not connected by local gates.
 
 \item The entanglement entropy of any state in $\Omega_d$ is 
 
 $$ 1- \log_2 \sqrt{\frac{(1+\sin 2 d)^{1+\sin 2 d}}{(1-\sin 2 d)^{-1+\sin 2 d}}}$$
 
 \item
 For any pair of states $\ket{\phi_1}$ and $\ket{\phi_2}$ in $ RQ_2$ there exists angles $\theta_0$, $\theta_1$,
 $\theta_2$ and $\theta_3$ such that the circuit 
 
 \vskip.5cm
 \centerline{
 \Qcircuit @C=1em @R=.7em {
  & \gate{R_y(\theta_1)}& \ctrl{1} &\gate{R_y(\theta_3)}  & \qw \\ 
  &  \gate{R_y(\theta_0)}&\ctrl{-1} & \gate{R_y(\theta_2)}& \qw
 }
 }
 
 sends $\ket{\phi_1}$ to $\ket{\phi_2}$.
 \vskip.2cm
 
 \item If $\ket{w}=w_1 \ket{00}+w_2 \ket{01}+w_3\ket{10}+w_4\ket{11}$ then the circuit 
 
 \vskip.5cm
 \centerline{
 \Qcircuit @C=1em @R=.7em {
 \lstick{\ket{0}} & \gate{R_y(\theta_1)}& \ctrl{1} & \qw                              & \qw \\ 
 \lstick{\ket{0}} &  \gate{R_y(\theta_0)}&\ctrl{-1} & \gate{R_y(\theta_2)}& \qw
 }
 }
 \vskip.2cm
 
 \begin{figure}[t]
  \begin{center}
  \includegraphics[width=.4\textwidth]{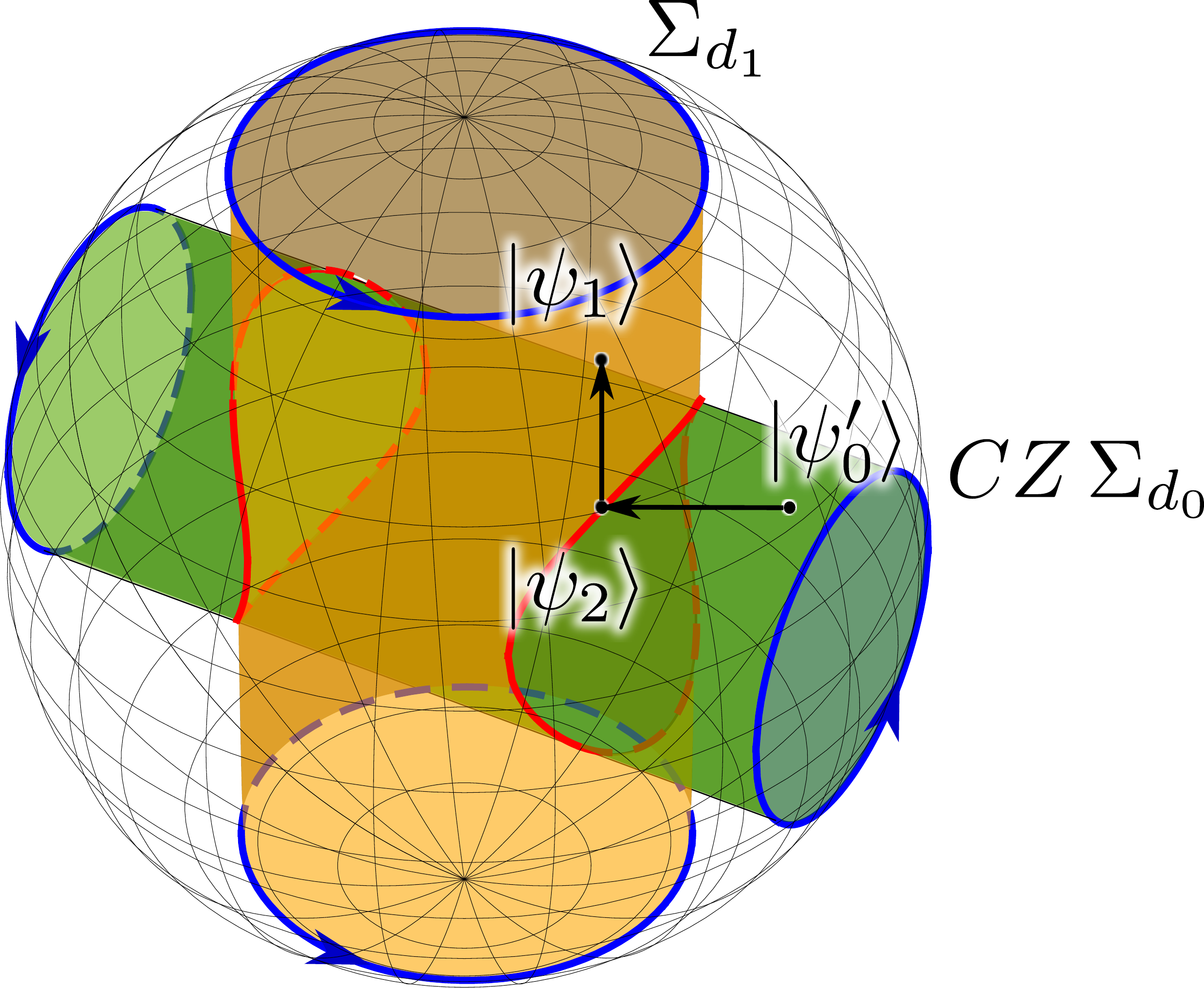}   
  \end{center}
  \caption{ 
  Geometrical proof that one CZ and local gates are enough to connect any 2-qubit states with real amplitudes. Here we show how to connect two states, a $\ket{\psi_0}$ in $\Sigma_{d_0}$ with a state $\ket{\psi_1}$ in $\Sigma_{d_1}$. Since the surface $\Sigma_{d_1}$ (orange) satisfy the equation  $x_1^2 +x_2^2=\sin^2(d_1)$, and the surface ${\rm CZ}\Sigma_{d_0}$ (green) satisfy the equation $x_2^2 +x_3^2=\sin^2(d)$ then these two surfaces intercept allowing to connect states from one surface to the other. In the figure, we depict the connection by local gates between $\ket{\psi_0'} = {\rm CZ} \ket{\psi_0}$ and $\ket{\psi_1}$ through $\ket{\psi_2}$ (a state in the interception).   
  }\label{fig:CZ}  
  \end{figure}

 with $\theta_0=Arg(w_1+i w_2)-Arg(w_3+i w_4)$, $\theta_1=2 \arccos(\sqrt{w_1^2+w_2^2})$ and $\theta_2=Arg(w_1+i w_2)+Arg(w_3+i w_4)$ prepares $\ket{w}$.

 \end{enumerate}

 \end{thm}

Proof:  Part (1) follows form the fact that all the elements in $E(\bm{v}_3,\bm{v}_4)$ are of the form  $(R_y(-2t)\otimes I_2) \ket{v_3}$ and all the element in $E(\bm{v}_1,\bm{v}_2)$ are of the form $(R_y(2 t)\otimes I_2) \ket{v_1}$. We also have that $(X\otimes I_2)\ket{v_1}=-\ket{v_4}$. Therefore all the elements in $E(\bm{v}_3,\bm{v}_4)\cap E(\bm{v}_1,\bm{v}_2)$ are part of the same orbit. We have that this union of two circle gives us the whole orbit because this set is closed under local gates with real amplitudes. Let us consider a positive $d<\pi/2$ and a state 
 
 $$\ket{\psi_0}=w_1 \ket{00}+w_2 \ket{01}+w_3\ket{10}+w_4\ket{11}\, \in \Sigma_d \ .$$
 
  Looking at the definition of $\Sigma_d(\bm{v}_3,\bm{v}_4)$ we conclude that there exists a pair of angles $a$ and $b$ such that 
 
 \begin{eqnarray*}
 \frac{w_1-w_4}{\sqrt{2}}&=&\sin(d) \cos(a) \qquad  \frac{w_2+w_3}{\sqrt{2}} =\sin(d) \sin(a) \\
 \frac{w_1+w_4}{\sqrt{2}}&=&  \cos(d)\cos(b) \qquad  \frac{w_2-w_3}{\sqrt{2}}=\cos(d)\sin(b)
 \end{eqnarray*}
 
 and therefore, 
 
 \begin{eqnarray*}
\ket{\psi_0} =   \big( \ (\cos (a) \sin (d)+\cos (b) \cos (d)) &\ket{00}& \\
 (\sin (a) \sin (d)+\sin (b) \cos (d))  & \ket{01}& +  \\
 (\sin (a) \sin (d)-\sin (b) \cos (d)) & \ket{10}&  +  \\
 (\cos (b) \cos (d)-\cos (a) \sin (d)) & \ket{11}& \big) / \sqrt{2}
  \end{eqnarray*}

 A direct verification shows that if we define the parametric surface $ \ket{ \phi(s,t)} $ by
 
\begin{eqnarray}
 R_y(2 s)\otimes R_y(2 t) \ket{ \psi_0} &=&  \phi_1\ket{00} +\phi_2\ket{01}+ \nonumber \\ 
                                        & &  \phi_3\ket{10}+\phi_4\ket{11} , 
\end{eqnarray}
with 
\begin{eqnarray}
\phi_1 &=& \big( \, 2 \sin (d) \cos (a+s+t)+\cos (b-d-s+t) \nonumber \\ 
       & & + \cos (b+d-s+t)\, \big) / (2\sqrt{2}) \nonumber \ ,\\
\phi_2 &=& \big( \, \cos (a-d+s+t)-\cos (a+d+s+t)+ \nonumber \\
       & & + 2 \cos (d) \sin (b-s+t)\, \big) / (2\sqrt{2}) \ , \nonumber \\
\phi_3 &=& \big( \, \cos (a-d+s+t)-\cos (a+d+s+t) \nonumber \\
       & & - 2 \cos (d) \sin (b-s+t) \, \big) / (2\sqrt{2}) \ ,  \nonumber \\
\phi_4 &=& \big( \,  -2 \sin (d) \cos (a+s+t)+\cos (b-d-s+t) \nonumber \\
       & & + \cos (b+d-s+t) \, \big) / (2\sqrt{2}) \ , \nonumber
\end{eqnarray}
 
 then, $\phi=(\phi_1,\phi_2,\phi_3,\phi_4)$ is an immersion because the vectors $\partial \phi /\partial s$ and $\partial \phi/\partial t$ are linear independent due to the fact that
 \begin{equation}
 \left| \frac{\partial \phi}{\partial s} \right|^2 \, 
 \left|\frac{\partial \phi}{\partial t}  \right|^2- 
 \left( \frac{\partial \phi}{\partial s} \cdot \frac{\partial \phi}{\partial t} \right)^2 = \sin^2(2d)>0\ . \nonumber
\end{equation}
 On the other hand, a direct verification shows that 
 
 \begin{eqnarray*}
 \frac{(\phi_1-\phi_4)^2}{2}+\frac{(\phi_2+\phi_3)^2}{2}=\sin^2(d) \\
 \frac{(\phi_1+\phi_4)^2}{2}+\frac{(\phi_2-\phi_3)^2}{2}=\cos^2(d),
 \end{eqnarray*}
 
 The last two equalities shows that $\phi(\mathbb{R}^2)$ is contained in $\Sigma_d(\bm{v}_3,\bm{v}_4)$. Since $\phi$ is double periodic (on its domain variables, $s$ and $t$) it defines a function on a torus. Being the torus a compact set and the map $\phi$ and immersion we conclude that $\phi$ is onto. A direct computation also shows that for any state $\phi(s,t)$, the state $(X\otimes I_2)\phi(s,t)$ is in $\Sigma_{\frac{\pi}{2}-d}(\bm{v}_3,\bm{v}_4)=\Sigma_d(\bm{v}_1,\bm{v}_2)$. We therefore conclude that any pair of states in $\Omega_d$ are connected by local gates.
 
 We continue the proof by pointing out that for any state $\phi(s,t)$ the eigenvalues of either one of the two trace matrices are 
 
 $$  \frac{1}{2}(1 - \sin (2d) ) \quad \hbox{and} \quad  \frac{1}{2} (1 + \sin (2 d)) \ . $$
 
 From the expression of the eigenvalues above we conclude that the entanglement entropy of the states on $\Omega_d$ is 
 \begin{equation}
  S(d)= 1- \log_2 \sqrt{\frac{(1+\sin 2 d)^{1+\sin 2 d}}{(1-\sin 2 d)^{-1+\sin 2 d}}}
 \end{equation}
 
 We can easily check that $\ket{00}$ is in $\Sigma_{\pi/4}$, therefore all elements in $\Sigma_{\pi/4}$ are not entangled states.
 Since the function $S(d)$ is one to one between $0$ and $\pi/4$ we conclude that the only unentangled states are those in $\Sigma_{\pi/4}$. Once again using the expression for the entanglement entropy,  we deduce that: two states in $\Omega_{d_1}$ and $\Omega_{d_2}$ respectively, are not connected by local gates if $d_1< d_2<\pi/4$. Recall that if two states are connected by local gates, then their entanglement entropy is the same. The previous arguments prove parts (2), (3) and (4). In order to prove part (5), we notice that the states of the form  ${\rm CZ} \phi(u,v)$, this is, the states
 
 $$\tilde{\phi}_1(u,v)\ket{00}+\tilde{\phi}_2(u,v)\ket{01}+\tilde{\phi}_3(u,v)\ket{10}+\tilde{\phi}_4(u,v)\ket{11}$$ 
 
 satisfy the equations $x_2^2+x_3^2=\sin^2(d)$ and $x_1^2+x_4^2=\cos^2(d)$. This is true because it can be easily checked that 
 
\begin{eqnarray*}
\frac{(\tilde{\phi}_1+\tilde{\phi}_4)^2}{2}+\frac{(\tilde{\phi}_2+\tilde{\phi}_3)^2}{2}=\sin^2(d) \\
 \frac{(\tilde{\phi}_1-\tilde{\phi}_4)^2}{2}+\frac{(\tilde{\phi}_2-\tilde{\phi}_3)^2}{2}=\cos^2(d)
 \end{eqnarray*} 
 Recall that the only difference between ${\rm CZ}\phi$ and $\phi$ is that the $\tilde{\phi}_4=-\phi_4$, the other coordinates with respect to the basis $\ket{00},\ket{01},\ket{10},\ket{11}$ are the same. Let us consider the state $\ket{\psi_0}$ in $\Sigma_d$ and another state $\ket{\psi_1}$ in $\Sigma_{d_1}$. We will see that the existence of the circuit using  only one CZ  gate and local gates is equivalent to the existence of point in the intersection $\Lambda={\rm CZ}\phi(\mathbb{R}^2)\cap\Sigma_{d_1}$. The reason is that once we have a state $\ket{\phi_2}$ in $\Lambda$, then, since $\ket{\phi_2}$  in  ${\rm CZ}\phi(\mathbb{R}^2)$ we have that $\ket{\phi_2}={\rm CZ}\ket{\phi_3}$ for some $\ket{\phi_3}$ in $\Sigma_d$ and since  $\ket{\phi_3}$ is in $\Sigma_d$ then $\ket{\phi_3}$  and $\ket{\phi_0}$ are connected by local gates. On the other hand since $\ket{\phi_2}$ is also on $\Sigma_{d_1}$, then $\ket{\phi_2}$ and $\ket{\phi_1}$ are connected by local gates. Therefore the states $\ket{\phi_0}$ and $\ket{\phi_1}$ are connected by   a circuit that only uses one CZ gate and local gates. The reason the surfaces ${\rm CZ}\phi(\mathbb{R}^2)$ and 
 the surface $\Sigma_{d_1}$ intersects is because they are two transversal tori. We can see this algebraically  by explicitly solving the equations
 \begin{eqnarray}\label{eq:eqsys}
  x_2^2+x_3^2 &= \sin^2(d), \quad x_1^2+x_4^2 &=\cos^2(d), \nonumber \\
  x_1^2+x_2^2 &= \sin^2(d_1),\quad  x_3^2+x_4^2 &=\cos^2(d_1) .  
\end{eqnarray}
 Since quantum circuits are reversibles, we can assume that $d>d_1$. A solution of the equation system \eqref{eq:eqsys} is given by 
 \begin{eqnarray}
 x_1 &=& 0, \nonumber \\ 
 x_2 &=& \sin(d_1),\nonumber \\
 x_3 &=& \sqrt{\sin^2(d)-\sin^2(d_1)}, \nonumber \\ 
 x_4 &=& \cos(d) \ .
 \end{eqnarray}
 
 For the proof of part (6), a direct calculation of the output state $\ket{w}$, from the quantum circuit given in part (6), yields  
 \begin{eqnarray}
  w_1 &=& \cos ( \theta_1 / 2 ) \cos(\theta_3) , \nonumber \\
  w_2 &=&  \cos ( \theta_1 /2 )\, \sin(\theta_3) , \nonumber \\
  w_3 &=& \sin ( \theta_1 / 2 ) , \nonumber \\
  w_4 &=& \sin(\theta_4)  , \nonumber 
\end{eqnarray}
the above result shows that $\theta_0=\theta_3-\theta_4$  and $\theta_2=\theta_3+\theta_4$.
 
\section{Conclusions} \label{s:conclusions}

In this work we described how the space of pure states of 2-qubit state with real amplitudes partitions when we group them by their entanglement.  We studied the geometry of each one of these groups when viewed as subsets of the 3-dimensional sphere $S^3$. Recall that the 2-qubit states with real amplitudes naturally identify with point in $S^3$.

The group that stands up for being different is the group of 2-qubits with maximal entanglement ($S=1$). These 2 quits form a pair of disjoint circle of radius 1 in $S^3$. This is the only group represented by a 1-dimensional manifold. Any other group is characterized by the property of being within the same distance to this pair of circles. They are either pair of tori when the entanglement entropy is positive but less than 1 and the group of 2-qubit states that have entanglement entropy zero form a torus.

We  show that any pair of 2-qubit states taken from the same group can be connected with $R_y$ and $X$ gates.

By considering the geometry of the action of the CZ gate on these groups, we provide a geometrical proof that every pair of 2-qubit states with real amplitudes can be connected with a circuit that contains only one CZ gate and  $R_y$ and $X$ gates. For the particular case when one of these 2-qubit states is the state $\ket{00}$, we explicitly provide the circuit.

Quantum states with real amplitudes find their applications in many domains: from stabilizers codes for quantum error correction \cite{Gottesman97} to probabilistic models to assist unsupervised machine learning tasks \cite{Perdomo2019,Leyton2019}. Understanding the structure of their Hilbert space and the different pathways towards their efficient preparation is one of the main motivation of this work. Natural extensions of this work include the extension to quantum states with a larger number of qubits.

\acknowledgements  

The authors would like to thank Eric Peterson, Raban Iten, Mario Krenn, Marcus P. da Silva, and Marcello Benedetti for useful discussions and feedback on an early draft of this manuscript. V. Leyton-Ortega was supported as part of the ASCR Quantum Testbed Pathfinder Program at Oak Ridge National
Laboratory under FWP \#ERKJ332.

\end{document}